\documentstyle[latexsym,prl,aps,t1enc]{revtex}

\newcommand{\rmi}{{\rm i}}

\newcommand{\deq}{\equiv}

\begin{document}

\draft
\title{Parametric excitation of plasma waves by gravitational radiation}
\author{Gert Brodin\thanks{
E-mail address: gert.brodin@physics.umu.se} and Mattias Marklund\thanks{%
E-mail address: mattias.marklund@physics.umu.se}}
\address{Department of Plasma Physics, Ume{\aa} University, S--901 87
Ume{\aa}, Sweden}
\maketitle

\begin{abstract}
We consider the parametric excitation of a Langmuir wave and an
electromagnetic wave by gravitational radiation, in a thin plasma on a
Minkowski background. We calculate the coupling coefficients starting from a
kinetic description, and the growth rate of the instability is found. The
Manley--Rowe relations are fulfilled only in the limit of a cold plasma. As a
consequence, it is generally difficult to view the process quantum
mechanically, i.e.\ as the decay of a graviton into a photon and a plasmon.
Finally we discuss the relevance of our investigation to realistic physical
situations.
\end{abstract}


The state of matter in regions where general relativistic treatments are
desirable is often the plasma state. Nevertheless, plasma physics and
general relativity are quite distinct areas of physics, and accordingly
there are comparatively few papers using a general relativistic framework
that include the electromagnetic forces in their treatments. 
However, there are a few exceptions to this rule, see for example Refs.\
1--5, where the plasma dynamics in a strong gravitational field is
considered \cite{Holcomb-Tajima,Daniel-Tajima},
relativistic transport equations for a
plasma are derived \cite{Georgiou}, a general relativistic version of the
Kelvin-Helmholtz theorem is derived \cite{Elsasser-Popel} and photon
acceleration by gravitational radiation is considered \cite
{Mendonca-Shukla-Bingham}. Omission of the electromagnetic effects for a
plasma subject to gravitational forces is possible because gravity alone do
not separate the charges, and in many cases the plasma can be treated as a
neutral fluid, in spite of its electromagnetic properties.

In this letter we will consider a simple model problem, that has two
interesting properties: Firstly, the process of investigation requires a
general relativistic description of the plasma dynamics to occur, and
secondly we demonstrate the possibility of charge separation induced by the
gravitational effects. We start from a monochromatic gravitational wave
propagating through a thin plasma superimposed on a flat background metric,
and consider the parametric excitation of a plasma wave and an
electromagnetic wave. The plasma wave - that has an increasing amplitude due
to the above mentioned mechanism - undergoes charge density oscillations.
Thus the parametric excitation process - which of course has no Newtonian
analog - shows that although no {\em direct} charge separation can be caused
by gravitational forces, {\em indirectly} charge density perturbations may
grow due to the gravitational and electromagnetic interaction. Theoretical
aspects of our problem, i.e.\ satisfaction of the Manley--Rowe relations and
the possibility of a quantum interpretation of the interaction, and the
relevance of our model problem to realistic physical situations  will be
discussed in the end of this letter.

We will refer to tensorial objects in terms of their
components in a coordinate basis. Note that as observers the natural way to
measure quantities is to project them onto a Lorentz frame $\{{\bf
  e}_{\hat a}, {\hat a} = 0, ..., 3\}$, i.e., a frame
in which the metric components become those of the Minkowski metric. The
frame components of the electric and magnetic fields $E^{\hat{a}}$ and $B^{%
\hat{a}}$ are defined by $F^{\hat{0}\hat{a}}=E^{\hat{a}}$, $F^{\hat{a}\hat{b}%
}=\delta _{\hat{c}\hat{d}}\epsilon ^{\hat{a}\hat{b}\hat{c}}B^{\hat{d}}$,
where $\epsilon ^{\hat{a}\hat{b}\hat{c}}$ are the components of the totally
skew tensor. Expanding ${\bf E}=E^{\hat{a}}{\bf e}_{\hat{a}}$ and
${\bf B}=B^{\hat{a}}{\bf e}_{\hat{a}}$ to first order in the metric's
coordinate 
components, we get that the electric field components in the coordinate
basis are the same as in the Lorentz frame. Furthermore, the magnetic
field components deviates only to first order in the metric components
from the Lorentz tetrad components, and it is therefore meaningful to
use the coordinate components in place of the tetrad components. This
would of course not be the case in a strong gravitational field such
as, say, the Kerr geometry.

The Vlasov equation for the distribution $f=f(x^{\mu },p^{a})$ (where $\mu
,\nu ,...=t,x,y,z$ and $a,b,...=x,y,z$) reads \cite{Ehlers} 
\begin{equation}
\frac{\partial f}{\partial t}+\frac{p^{a}}{p^{t}}\frac{\partial f}{\partial
x^{a}}+\left( \frac{q}{p^{t}}g_{\mu\nu}F^{a\mu }p^{\nu}-%
{\cal {G}}^{a}\right) \frac{\partial f}{\partial p^{a}}=0\ ,
\label{eq:vlasov}
\end{equation}
where ${\cal G}^{a}\equiv\Gamma _{\mu \nu }^{a}p^{\mu }p^{\nu }/p^{t}$, and
this is coupled to Maxwell's equations 
\begin{mathletters}
\label{eq:Maxwell}
\begin{eqnarray}
  F^{\mu \nu }\!_{;\nu } = \mu _{0}j^{\mu }{\equiv}
  \sum_{{\rm p.s.}}
  \mu_{0}q\int f(x^{\nu },p^{a})p^{\mu
    }\frac{|g|^{1/2}}{|p_{t}|}d^{3}p \ , 
\label{eq:Maxwell1} \\
F_{[\mu \nu ,\sigma ]}  = 0 \ ,  \label{eq:Maxwell2}
\end{eqnarray}
where ${\rm p.s.}$\ stands for particle species. Here we have introduced the
invariant measure $(|g|^{1/2}/|p_{t}|)d^{3}p$ on the surface in momentum
space where $g_{\mu \nu }p^{\mu }p^{\nu }=-m^{2}$, $m$ being the mass of the
particle in question, and we use $g$ to denote the determinant of the
metric. The gravitational field, represented by the Christoffel
symbols $\Gamma_{\nu \sigma }^{\mu }$ in the Vlasov equation, is
assumed to be generated by some outside source. The plasma itself is
assumed to generate a much weaker gravitational field.

Next we assume the presence of a small amplitude gravitational wave,
and we use the transverse traceless gauge. The line element then takes
the form \cite{MTW} 
\end{mathletters}
\begin{equation}
  ds^{2} = -dt^{2} + (1 + h_{+})\,dx^{2} + (1 - h_{+})\,dy^{2} +
  2h_{\times}\,dx\,dy + dz^{2}\ ,
\end{equation}
where $h_{+}=h_{+}(u)$, $h_{\times }=h_{\times }(u)$, with
$u{\equiv}z-t$, and $|h_{+}|,$ $|h_{\times }|\ll 1$.
The mass of a particle with momentum $p^{\mu }$ is defined as
$m{\equiv }|g_{\mu \nu }p^{\mu }p^{\nu }|^{1/2}$. From this, $p^{t}$
may be expressed in terms of $m$ and the 3-momentum $p^{a}$: 
\begin{equation}
p^{t}=\left\{ m^{2}+\delta _{ab}p^{a}p^{b}+h_{+}\left[
(p^{x})^{2}-(p^{y})^{2}\right] +2h_{\times }p^{x}p^{y}\right\} ^{1/2}\ .
\label{eq:energy}
\end{equation}
To first order in $h_{+}$, $h_{\times }$, the Christoffel symbols read
\begin{mathletters}
\label{eq:Christoffel}
\begin{eqnarray}
&&-\Gamma _{xx}^{t}=\Gamma _{yy}^{t}=-\Gamma _{tx}^{x}=\Gamma
_{xz}^{x}=\Gamma _{ty}^{y}=-\Gamma _{yz}^{y}=-\Gamma _{xx}^{z}=\Gamma
_{yy}^{z}=\textstyle\frac{1}{2}\dot{h}_{+}\ ,  \label{eq:Christoffel1} \\
&&-\Gamma _{xy}^{t}=-\Gamma _{ty}^{x}=\Gamma _{yz}^{x}=-\Gamma
_{tx}^{y}=\Gamma _{xz}^{y}=-\Gamma _{xy}^{z}=\textstyle\frac{1}{2}\dot{h}%
_{\times }\ ,  \label{eq:Christoffel2}
\end{eqnarray}
\end{mathletters}
where $\dot{h}{\equiv }dh/du$.
Using the Christoffel symbols (\ref{eq:Christoffel}) of the gravitational
wave, the Vlasov equation (\ref{eq:vlasov}) for the unperturbed ($F^{\mu\nu}
= 0$) distribution $f=f(t,x^a,p^b)$ becomes 
\begin{equation}
\frac{\partial f}{\partial t}+\frac{p^{a}}{p^{t}}\frac{\partial f}{\partial
x^{a}}-{\cal G}^{a}\frac{\partial f}{\partial p^{a}}=0\ ,  \label{eq:vlasov2}
\end{equation}
where 
\begin{eqnarray*}
&&{\cal G}^{x}{\equiv} \left( -1+\frac{p^{z}}{p^{t}}\right) \left( \dot{h}%
_{+}p^{x}+\dot{h}_{\times }p^{y}\right) \ , \\
&&{\cal G}^{y}{\equiv} \left( -1+\frac{p^{z}}{p^{t}}\right) \left( \dot{h}%
_{\times }p^{x}-\dot{h}_{+}p^{y}\right) \ , \\
&&{\cal G}^{z}{\equiv} {\frac{1}{2p^{t}}}\left( \gamma _{AB}p^{A}p^{B}\dot{h}%
_{+}-2p^{x}p^{y}\dot{h}_{\times} \right) \ ,
\end{eqnarray*}
and $(\gamma _{AB}) = {\rm diag}(-1,1)$, $A,B=x,y$.
Expanding $p^{t}$ according to $p^{t} \approx p_{(0)}^{t} - {\cal F}$, where 
${\cal F} \equiv \left( h_+\gamma_{AB}p^Ap^B -
2h_{\times}p^xp^y\right)/(2p^t_{(0)})$, and $p_{(0)}^{t} = (m^{2}+\delta
_{ab}p^{a}p^{b})^{1/2}$, the Vlasov equation becomes 
\begin{equation}
\frac{\partial f}{\partial t} + \left( 1 + \frac{{\cal F}}{p^t_{(0)}}
\right) \frac{p^a}{p^t_{(0)}}\frac{\partial f}{\partial x^a} - {\cal G}^{a}%
\frac{\partial f}{\partial p^{a}}=0\ ,  \label{Vlasov3}
\end{equation}

For simplicity we let the gravitational wave be monochromatic:  
\begin{equation}
  h = \tilde{h}\exp \left[ {\rm i}(k_{0}z - \omega _{0}t)\right] +
  {\rm c.c.} \ , 
\end{equation}
with the dispersion relation $k_{0}=\omega_0$.
We divide $f$ according to $f = f_{{\rm SJ}}(p^{a}) + f_{{\rm
    g}}(t,z,p^{a})$
where the equilibrium distribution $f_{{\rm SJ}}$ is taken to be the
Synge--J{\"u}ttner distribution \cite{Liboff}, and $f_{{\rm g}}=\widetilde{%
f_{{\rm g}}}\exp \left[ {\rm i}(k_{0}z-\omega _{0}t)\right] $ is the
perturbation induced by the gravitational wave - which is assumed to fulfill 
$f_{{\rm g}}\ll f_{{\rm SJ}}$. From Eq.\ (\ref{Vlasov3}) we find 
\begin{equation}
\widetilde{f_{{\rm g}}}=-\frac{{\cal G}^{a}(\partial f_{{\rm SJ}}/\partial
p^{a})}{{\rm i}\left( \omega _{0}-k_{0}p^{z}/p_{(0)}^{t}\right) }
\end{equation}
to first order in the amplitude.

Next we assume the presence of electromagnetic perturbations with frequency
and wavenumber $(\omega _{1},{\bf k}_{1})$ as well as electrostatic
perturbations $(\omega _{2},{\bf k}_{2}).$ The frequencies and wavenumbers
are taken to satisfy the resonance conditions $\omega _{0}=\omega
_{1}+\omega _{2}$ and ${\bf k}_{0}{\bf =k}_{1}{\bf +k}_{2}.$ The vectors $%
{\bf k}_{1}$ and ${\bf k}_{2}$ span a plane, and we chose the $y$-axis to be
perpendicular to this plane, i.e.\ ${\bf k}_{1}=k_{1}^{x}\widehat{{\bf x}}$ $%
+k_{1}^{z}\widehat{{\bf z}}$ and ${\bf k}_{2}=k_{2}^{x}\widehat{{\bf x}}$ $%
+k_{2}^{z}\widehat{{\bf z}}.$ Including the electromagnetic field, expanding
Eq.\ (\ref{Vlasov3}) to first order in $h$ and writing all terms proportional
to $h$ on the right hand side Eq.\ (\ref{Vlasov3}) becomes 
\begin{eqnarray}
&&\frac{\partial f_{{\rm ng}}}{\partial t}+\frac{p^{a}}{p_{(0)}^{t}}\frac{%
\partial f_{{\rm ng}}}{\partial x^{a}}+q\left( -F^{at}+\frac{1}{p_{(0)}^{t}}%
\delta _{bc}F^{ac}p^{b}\right) \frac{\partial f_{{\rm ng}}}{\partial p^{a}} 
\nonumber \\
&=&-\frac{{\cal F}}{(p_{(0)}^{t})^{2}}p^{a}\frac{\partial f_{{\rm ng}}}{%
\partial x^{a}}+{\cal G}^{a}\frac{\partial f_{{\rm ng}}}{\partial p^{a}}%
-q\left( -F^{at}+\frac{1}{p_{(0)}^{t}}\delta _{bc}F^{ac}p^{b}\right) \frac{%
\partial f_{{\rm g}}}{\partial p^{a}}  \nonumber \\
&&-\left\{ \frac{q}{p_{(0)}^{t}}\left[ -h_{+}\gamma
_{BC}F^{aC}p^{B}+h_{\times }\left( F^{ax}p^{y}+F^{ay}p^{x}\right) \right]
\right\} \frac{\partial f_{{\rm SJ}}}{\partial p^{a}}  \label{Vlasov4}
\end{eqnarray}
where we have introduced $f_{{\rm ng}}$ defined by $f_{{\rm
    ng}}=f-f_{{\rm g}}$.
Furthermore, we need two components of Faraday's law. Expanding 
Eq.\ (\ref{eq:Maxwell2}) to first order in the amplitude $h$, we
obtain  
\begin{mathletters}
\label{eq:Maxwell3}
\begin{eqnarray}
  \partial_{t}F^{xy} - \partial_yF^{tx} + \partial_xF^{ty}
  = h_{\times}\left( -\partial_{x}F^{tx} + \partial_yF^{ty} \right)
  + h_+\left( \partial_yF^{tx} + \partial_xF^{ty} \right)
    \ ,  \label{Faraday_z} \\
  \partial_tF^{yz} - \partial_zF^{ty} + \partial_yF^{tz}  
  = h_{\times}\left( \partial_zF^{tx} - \partial_tF^{xz} \right)
  + h_+\left( \partial_zF^{ty} + \partial_tF^{yz} \right)
   \nonumber \\
    \qquad \qquad \qquad + \dot{h}_{\times}\left( F^{tx} +
  F^{xz}\right) - \dot{h}_{+}\left( F^{ty} + F^{yz}\right)  \
  ,\label{Faraday_x} 
\end{eqnarray}
\end{mathletters}

In what follows, we will put $h_{+}=0$, i.e., we choose a specific
polarization of the gravitational wave. It turns out that such a
gravitational wave does not couple to a linearly polarized electromagnetic
wave with magnetic field in the $y$-direction. Thus we assume the
electromagnetic wave to have the opposite polarization, i.e.\ $%
F^{yt}\!,F^{xz}\!$ and $F^{yz}\!$ are the only electromagnetic components to
be different from zero in the linear approximation. Similarly $F^{zt}\!$ and 
$F^{xt}\!$ are the only nonzero electrostatic components of the field
tensor. We then divide $f_{{\rm ng}}$ according to $f_{{\rm ng}}=f_{{\rm SJ}%
}(p^{a})+\widetilde{f_{{\rm em}}}(t,p^{a})\exp \left[ {\rm i}%
(k_{1}^{x}x+k_{1}^{z}z-\omega _{1}t)\right] +\widetilde{f_{{\rm es}}}%
(t,p^{a})\exp \left[ {\rm i}(k_{2}^{x}x+k_{2}^{z}z-\omega _{2}t)\right] $
where the time dependence of the amplitudes, which is due to the parametric
interaction with the gravitational wave, is assumed to be slow, such
that $\partial /\partial t\ll \omega .$ In the linear approximation (i.e.\ no
gravitational coupling) the slow time dependence vanishes, and Eq.\ (\ref
{Vlasov4}) gives 
\begin{mathletters}
\begin{eqnarray}
\widetilde{f_{{\rm em}}^{{\rm l}}} &=&-\frac{q\widetilde{F}^{yt}}{{\rm i}%
\widehat{\omega }_{1}}\frac{\partial f_{{\rm SJ}}}{\partial p^{y}}\ ,
\label{flem} \\
\widetilde{f_{{\rm es}}^{{\rm l}}} &=&-\frac{q\widetilde{F}^{at}}{{\rm i}%
\widehat{\omega }_{2}}\frac{\partial f_{{\rm SJ}}}{\partial p^{a}}
\label{fles}
\end{eqnarray}
for the electromagnetic and electrostatic perturbations respectively. Here
we have $\widehat{\omega }_{i}{\equiv }\omega _{i}-{\bf k}_{i}{\bf \cdot }%
{\bf p}/p_{(0)}^{t}$, where the scalar product is defined as
${\bf A\cdot B}{\equiv }\delta _{ab}A^{a}B^{b}$.
Considering the part of Eq.\ (\ref{Vlasov4}) varying as $\exp \left[ {\rm i}%
(k_{2}^{x}x+k_{2}^{z}z-\omega _{2}t)\right] $ using linear approximations
for the factors on the right hand side we obtain $\widetilde{f_{{\rm es}}}=%
\widetilde{f_{{\rm es}}^{{\rm l}}}+\widetilde{f_{{\rm es}}^{{\rm nl}}}$. The
linear contribution is given by (\ref{fles}), by replacing $\omega _{2}$
by $\omega _{2{\rm op}}{\equiv }\omega _{2} + {\rmi}\partial /\partial
t$ (since $%
\partial /\partial t\ll \omega _{2}$, division by $\omega _{2{\rm
    op}}$ can be calculated by a first order Taylor expansion) and the
nonlinear contribution is found to be $\widetilde{f_{{\rm es}}^{{\rm
    nl}}}=C_{{\rm es}}\widetilde{h}_{\times
    }{\widetilde{F}^{yt}\!}^{*}$ where  
\end{mathletters}
\begin{eqnarray}
C_{{\rm es}} &=&\frac{q}{{\rm i}\widehat{\omega }_{2}}\left\{ \frac{%
p^{x}p^{y}{\bf k}_{1}\cdot {\bf p}}{(p_{(0)}^{t})^{3}\widehat{\omega }_{1}}%
\frac{\partial f_{{\rm SJ}}}{\partial p^{y}}+\omega _{0}{\cal G}_{{\rm op}%
}\left( \frac{1}{\widehat{\omega }_{1}}\frac{\partial f_{{\rm SJ}}}{\partial
p^{y}}\right) \right.   \nonumber \\
&&+\left[ \left( 1-\frac{1}{p_{(0)}^{t}}\left( \frac{{\bf p\cdot k}_{1}}{%
\omega _{1}}\right) \right) \frac{\partial }{\partial p^{y}}+\frac{p^{y}}{%
\omega _{1}p_{(0)}^{t}}\left( k_{1}^{x}\frac{\partial }{\partial p^{x}}%
+k_{1}^{z}\frac{\partial }{\partial p^{z}}\right) \right] \left. \left[ 
\frac{\omega _{0}}{\widehat{\omega }_{0}}\left( {\cal G}_{{\rm op}}f_{{\rm SJ%
}}\right) \right] +\frac{k_{1}^{z}p^{x}}{p_{(0)}^{t}\omega _{1}}\frac{%
\partial f_{{\rm SJ}}}{\partial p^{z}}\right\}   \label{fes_nl2}
\end{eqnarray}
and 
$
{\cal G}_{{\rm op}}{\equiv }(1-p^{z}/p_{(0)}^{t})(p^{y}\partial
/\partial p^{x}+p^{x}\partial /\partial
p^{y})+(p^{x}p^{y}/p_{(0)}^{t})\partial/\partial p^{z}\ .
$
Combining Eqs.\ (\ref{fes_nl2}) and (\ref{eq:Maxwell1}) we then find  
\begin{equation}
\varepsilon _{L}(\omega _{2{\rm op}},k_{2})\widetilde{F}^{es}\!=\frac{\mu
_{0}q\widetilde{h}_{\times }{\widetilde{F}^{yt}\!}^{*}}{{\rm i}k_{2}}\sum_{%
{\rm p.s.}}\int C_{{\rm es}}{d^{3}p.}  \label{estat_1}
\end{equation}
where the longitudinal dielectric permittivity $\varepsilon _{L}$ is given
by \cite{Brodin-Stenflo} 
\begin{equation}
\varepsilon _{L}=1+\frac{\mu _{0}q}{k_{2}}\sum_{{\rm p.s.}}\int \frac{1}{%
\widehat{\omega }_{2{\rm op}}}\frac{\partial f_{{\rm SJ}}}{\partial p}{%
d^{3}p,}  \label{eps_L}
\end{equation}
$p=[(p^{x})^{2}+(p^{y})^{2}+(p^{z})^{2}]^{1/2}$, $%
k_{2}=[(k_{2}^{x})^{2}+(k_{2}^{z})^{2}]^{1/2}$, and $\widetilde{F}^{es}=[(%
\widetilde{F}^{xt})^{2}+(\widetilde{F}^{zt})^{2}]^{1/2}$. In deriving
Eqs.\ (\ref{estat_1}) and (\ref{eps_L}) we have used that the
electrostatic mode is longitudinal,
i.e. $\widetilde{F}^{xt}/\widetilde{F}^{zt}=k_{2}^{x}/k_{2}^{z}$. 

Next we turn to the part of (\ref{Vlasov4}) varying as $\exp \left[
  {\rm i}(k_{1}^{x}x+k_{1}^{z}z-\omega _{1}t)\right] $. Again using
linear approximations for the right hand side we obtain
  $\widetilde{f_{{\rm em}}} = \widetilde{f_{{\rm em}}^{{\rm
  l}}}+\widetilde{f_{{\rm em}}^{{\rm nl}}}$ 
where $\widetilde{f_{{\rm em}}^{{\rm l}}}$ is given by (\ref{flem})
(naturally by replacing $\omega _{1}$ with $\omega _{1{\rm op}}$), and $%
\widetilde{f_{{\rm em}}^{{\rm nl}}}$ is found to be
$\widetilde{f_{{\rm em}}^{{\rm nl}}}=C_{{\rm em}}\widetilde{h}_{\times
  }{\widetilde{F}^{es}\!}^{*}$ where 
\begin{eqnarray}
C_{{\rm em}} &=&\frac{q}{{\rm i}\widehat{\omega }_{1}}\left\{ \omega _{0}%
{\cal G}_{{\rm op}}\left( \frac{\widehat{k}_{2}^{x}}{\widehat{\omega }_{2}}%
\frac{\partial f_{{\rm SJ}}}{\partial p^{x}}+\frac{\widehat{k}_{2}^{z}}{%
\widehat{\omega }_{2}}\frac{\partial f_{{\rm SJ}}}{\partial p^{z}}\right)
\right.   \nonumber \\
&&-\left( \widehat{k}_{2}^{x}\frac{\partial }{\partial p^{x}}+\widehat{k}%
_{2}^{z}\frac{\partial }{\partial p^{z}}\right) \left[ \frac{\omega _{0}}{%
\widehat{\omega }_{0}}{\cal G}_{{\rm op}}f_{{\rm SJ}}\right] \left. -\frac{%
p^{x}p^{y}{\bf k}_{2}\cdot {\bf p}}{(p_{(0)}^{t})^{3}}\left( \frac{\widehat{k%
}_{2}^{x}}{\widehat{\omega }_{2}}\frac{\partial f_{{\rm SJ}}}{\partial p^{x}}%
+\frac{\widehat{k}_{2}^{z}}{\widehat{\omega }_{2}}\frac{\partial f_{{\rm SJ}}%
}{\partial p^{z}}\right) \right\} .  \label{fem_nl}
\end{eqnarray}
Eliminating linear terms proportional to $F^{yz}$ and $F^{xy}$ with the help
of Eqs. (\ref{Faraday_x}) and  (\ref{Faraday_z}), Eq.\ (\ref{eq:Maxwell1})
together with (\ref{fem_nl}) gives 
\begin{eqnarray}
D(\omega _{1{\rm op}},|{\bf k}_{1}|)\widetilde{F}^{yt}\! &&{\equiv }\left(
\varepsilon _{T}(\omega _{1{\rm op}},|{\bf k}_{1}|)-\frac{{\bf k}%
_{1}^{2}c^{2}}{\omega _{1{\rm op}}^{2}}\right) \widetilde{F}^{yt}\!
\nonumber \\
&=&\left\{ \frac{{\rm i}\mu _{0}q}{\omega _{1}}\int \left[ \frac{p^{y}}{%
p_{(0)}^{t}}\left( C_{{\rm em}}+\frac{qp^{x}p^{y}}{{\rm i}(p_{(0)}^{t})^{2}}%
\left( \frac{k_{2}^{x}}{k_{2}\widehat{\omega }_{2}}\frac{\partial f_{{\rm SJ}%
}}{\partial p^{x}}+\frac{k_{2}^{z}}{k_{2}\widehat{\omega }_{2}}\frac{%
\partial f_{{\rm SJ}}}{\partial p^{z}}\right) \right) \right] d^{3}p+\frac{%
k_{2x}k_{1}^{2}}{k_{2}\omega _{1}^{2}}\right\} \widetilde{h}_{\times }{%
\widetilde{F}^{es}\!}^{*}  \label{emagn_1} 
\end{eqnarray}
where the transverse dielectric permittivity $\varepsilon _{T}$ is given by 
\cite{Brodin-Stenflo} 
\begin{equation}
\varepsilon _{T}=1+\frac{\mu _{0}q^{2}}{\omega _{1}}\sum_{{\rm p.s.}}\int 
\frac{p^{y}}{\widehat{\omega }_{1}}\frac{\partial f_{{\rm SJ}}}{\partial
p^{y}}{d^{3}p.}  \label{eps_T}
\end{equation}
The operators $\varepsilon _{L}(\omega _{2{\rm op}},k_{2})$ and $D(\omega _{1%
{\rm op}},|{\bf k}_{1}|)$ can be expanded according to $\varepsilon
_{L}(\omega _{2{\rm op}},k_{2})=\varepsilon _{L}(\omega
_{2},k_{2})+(\partial \varepsilon _{L}/\partial \omega _{2}){\rm i}\partial
/\partial t$ and $D(\omega _{1{\rm op}},|{\bf k}_{1}|)=D(\omega _{1},|{\bf k}%
_{1}|)+(\partial D/\partial \omega _{1}){\rm i}\partial /\partial t.$
Assuming that the dispersions relations for the Langmuir wave and the
electromagnetic wave are exactly fulfilled, we let $\varepsilon _{L}(\omega
_{2},k_{2})=D(\omega _{1},|{\bf k}_{1})=0$. Equations (\ref{estat_1})
and (\ref{emagn_1}) are then written 
\begin{mathletters}
\begin{eqnarray}
\frac{\partial {\widetilde{F}^{es}\!}}{\partial t} &=&C_{1}\frac{\widetilde{h%
}_{\times }{\widetilde{F}^{yt}\!}^{*}}{(\partial \varepsilon _{L}/\partial
\omega _{2})}\ ,  \label{estat_2} \\
\frac{\partial {\widetilde{F}^{yt}\!}}{\partial t} &=&C_{2}\frac{\widetilde{h%
}_{\times }{\widetilde{F}^{es}\!}^{*}}{(\partial D/\partial \omega _{1})}
\label{emagn_2}
\end{eqnarray}
where the (constant) coupling coefficients $C_{1}$ and $C_{2}$ are trivially
found from (\ref{estat_1}) and (\ref{emagn_1}). As can be seen, generally
the expressions for the coupling coefficients are very complicated. In order
to get transparent formulas, we present the result for a cold electron
plasma with immobile ions constituting a neutralizing background. Letting $%
f_{{\rm SJ}}\rightarrow \delta ^{3}({\bf p})$ the coupling coefficients
become $C_{1}=C_{2} \deq C = -{\rmi}k_{2x}\omega _{0}/k_{2}\omega
_{1}.$ Using these expressions and combining (\ref{estat_2}) and
(\ref{emagn_2}) the growth rate $\gamma$ for the Langmuir wave and the
electromagnetic wave is given by 
\end{mathletters}
\begin{equation}
  \gamma^2 = \left(\frac{k_{2x}\omega _{0}}{k_{2}\omega
      _{1}}\right)^2\frac{\omega_1\omega_2}{4}|\widetilde{h}_{\times }|^2 \ .
  \label{growth}
\end{equation}
Since we have not included dissipation of the wave
modes, the threshold value for the instability is zero. However, it is
straightforward \cite{Weiland-Wilhelmsson} to calculate the threshold by
including appropriate damping mechanisms for the decay products.

The fact that the same coupling coefficient $C$ appears in Eqs.\ (%
\ref{estat_2}) and (\ref{emagn_2}) for a cold plasma means that the
Manley--Rowe relations \cite{Weiland-Wilhelmsson} are fulfilled for that
case. These relations usually follows from an underlying Hamiltonian
structure of the governing equations, and assures that each of the decay
products takes energy from the pump wave in direct proportion to their
respective frequencies. Thus fulfillment of the Manley--Rowe relation means
that the parametric process can be interpreted quantum mechanically - i.e.
we can think of the interaction as the decay of a graviton with energy $%
\hbar \omega _{0}$ into a photon with energy $\hbar \omega _{1}$ and a
plasmon with energy $\hbar \omega _{2}$. However, we stress that this
interpretation is not always possible. When thermal effects are taken into
account, the coefficient $C_{1}$ in (\ref{estat_2}) is generally different
from the coefficient $C_{2}$ in (\ref{emagn_2}). Thus we obtain the
rather surprising result that the simple graviton
interpretation (i.e.\ that the gravitational wave is built up of quanta with
energy $\hbar \omega _{0}$ ) of the gravitational field is not always
applicable. 

To our knowledge, our set of governing Eqs.\
(\ref{eq:vlasov})--(\ref{eq:Maxwell}), have not been used to
consider specific physical processes
previously. Probably this is because the system has a very high complexity -
it is nonlinear and has seven independent variables - but, on the other
hand, simpler models for the plasma cannot deal with thermal effects
properly. As a starting point we have addressed a simple ''model problem'',
to reduce the complexity of our governing equations. The main conclusion
that can be drawn from our calculations are two: Firstly, wave-wave
interactions can lead to energy transfer from gravitational to
electromagnetic degrees of freedom, and vice versa.
Secondly, charge separation and
corresponding electrostatic fields may occur as a result of such
interactions. However, the idealizations made in our calculations make it
difficult to make specific physical predictions. For example, binary systems
can produce monochromatic gravitational radiation, but the amplitude is
probably to low to overcome the threshold value caused by plasma
inhomogeneities \cite{Liu}, except possibly in the very last collapse phase
when the system produces an incoherent broadband gravitational pulse.
Similarly, supernova events may produce high enough gravitational wave
amplitudes for significant energy transfer to electromagnetic degrees of
freedom to occur, but in most cases we expect the gravitational
radiation to be very incoherent and broadband (there are
calculations predicting highly monochromatic emissions of
gravitational waves for certain types of neutron star formation
\cite{Ipser-Managan}). Some estimates of the effects due to scattering
of gravitational waves in supernova has been presented by Bingham {\it
  et al.}\ \cite{Bingham}. 
One may also consider the
formation of black holes as a source of gravitational radiation,
especially the non-axisymmetric collapse, but - again - this process
gives a pulse 
rather than a monochromatic wave. (See the review
by Thorne \cite{Thorne} for characteristic values of the amplitudes,
frequencies, and duration times for the processes discussed above.)

The parametric process considered in this letter can be thought of as
a gravitational analog of Raman scattering, where the electromagnetic
pump wave has been replaced by a 
gravitational wave. Similarly, we can imagine that a gravitational analog of
Compton scattering may occur. Such a process requires a kinetic treatment,
and could be described within our formalism without very much extra
difficulty. Gravitational waves may also be subject to four wave
processes such as modulational instabilities. Furthermore, a
gravitational radiation background similar to the microwave background
is assumed to exist as a relic from big bang \cite{Peebles}. In the
early universe the intensity of such radiation could be very high, and
- if the magnitude of the gravitational waves are sufficient -
processes such as those mentioned above could be of importance for the
structure formation in the universe. It should be stated though that
estimates of the amplitude of such primordial gravitational waves are
very difficult to make, much depending on the insufficient data
concerning the early universe. Also, much of the simplifying
assumption made in this letter must be relaxed before such issues
can be studied seriously.

\end{document}